\documentclass{article}

\usepackage{nips15submit_e,times}
\usepackage{url}

\usepackage[latin1]{inputenc}
\usepackage[english]{babel}
\usepackage[comma,authoryear]{natbib}
\usepackage{amsmath}
\usepackage{amsthm}
\usepackage{amsfonts}
\usepackage{amssymb}
\usepackage{graphicx} \graphicspath{{./figures/}}
\usepackage{epsfig}
\usepackage{subfig}
\usepackage{color}
\usepackage{todonotes}
\usepackage{endnotes}
\usepackage{titlesec}
\usepackage[titletoc,title]{appendix}

\newtheorem*{trm:bald_optimum_h}{Theorem}

\newcommand{\dB}{\ensuremath{\mathrm{dB}}}
\newcommand{\dBHL}{\ensuremath{\mathrm{dB HL}}}
\newcommand{\Hz}{\ensuremath{\mathrm{Hz}}}
\newcommand{\kHz}{\ensuremath{\mathrm{kHz}}}
\newcommand{\bark}{\ensuremath{\text{bark}}}
\newcommand{\+}[1]{\ensuremath{\mathbf{#1}}}
\newcommand{\Ent}{\ensuremath{\mathrm{H}}}
\newcommand{\ent}{\ensuremath{\mathrm{h}}}

\title{A Bayesian binary classification approach to pure tone audiometry}
\author{
  Marco Cox\\
  Eindhoven University of Technology\\
  Eindhoven, The Netherlands\\
  \texttt{m.g.h.cox@tue.nl}
  \AND
  Bert de Vries\\
  GN ReSound and Eindhoven University of Technology\\
  Eindhoven, The Netherlands\\
  \texttt{bdevries@ieee.org}
}

\nipsfinalcopy

\begin{document}

\maketitle

\begin{abstract}
The pure tone hearing threshold is usually estimated from responses to stimuli at a set of standard frequencies. This paper describes a probabilistic approach to the estimation problem in which the hearing threshold is modelled as a smooth continuous function of frequency using a Gaussian process. This allows sampling at any frequency and reduces the number of required measurements. The Gaussian process is combined with a probabilistic response model to account for uncertainty in the responses. The resulting full model can be interpreted as a two-dimensional binary classifier for stimuli, and provides uncertainty bands on the estimated threshold curve. The optimal next stimulus is determined based on an information theoretic criterion. This leads to a robust adaptive estimation method that can be applied to fully automate the hearing threshold estimation process.
\end{abstract}

\section{Introduction}
\label{sec:introduction}
The most common and basic way to quantify hearing impairment is to estimate the pure tone hearing threshold (HT), which is called \textit{pure tone audiometry} (PTA) \citep{yost_fundamentals_1994}. The conventional way to estimate the pure tone HT of a person is to incrementally approximate the lowest tone intensity that can still be perceived at a set of standard frequencies ranging from $250~\Hz$ to $8~\kHz$ using a staircase ``up $5~\dB$ - down $10~\dB$'' approach \citep{carhart1959preferred}. This simple approach however suffers from a number of drawbacks:
\begin{itemize}
\item It does not provide an uncertainty measure for the estimated thresholds, making it hard to define an objective stopping criterion.
\item It is difficult to exploit potential prior knowledge about the threshold curve that is being estimated.
\item It does not use all information available in the data by assuming that the hearing thresholds at different frequencies are uncorrelated. This leads to redundant measurements being required to achieve the desired accuracy.
\item Restricting the stimuli to a fixed set of standard frequencies requires additional assumptions to estimate the threshold at other frequencies. The most common solution to this is to assume piecewise linearity of the HT curve on a semi-logarithmic frequency scale, for example in an audiogram. An audiogram depicts the difference between a particular HT curve and the HT curve of a normal-hearing person.
\end{itemize}

\begin{figure}[!ht]
\centering
\includegraphics[width=.8\textwidth]{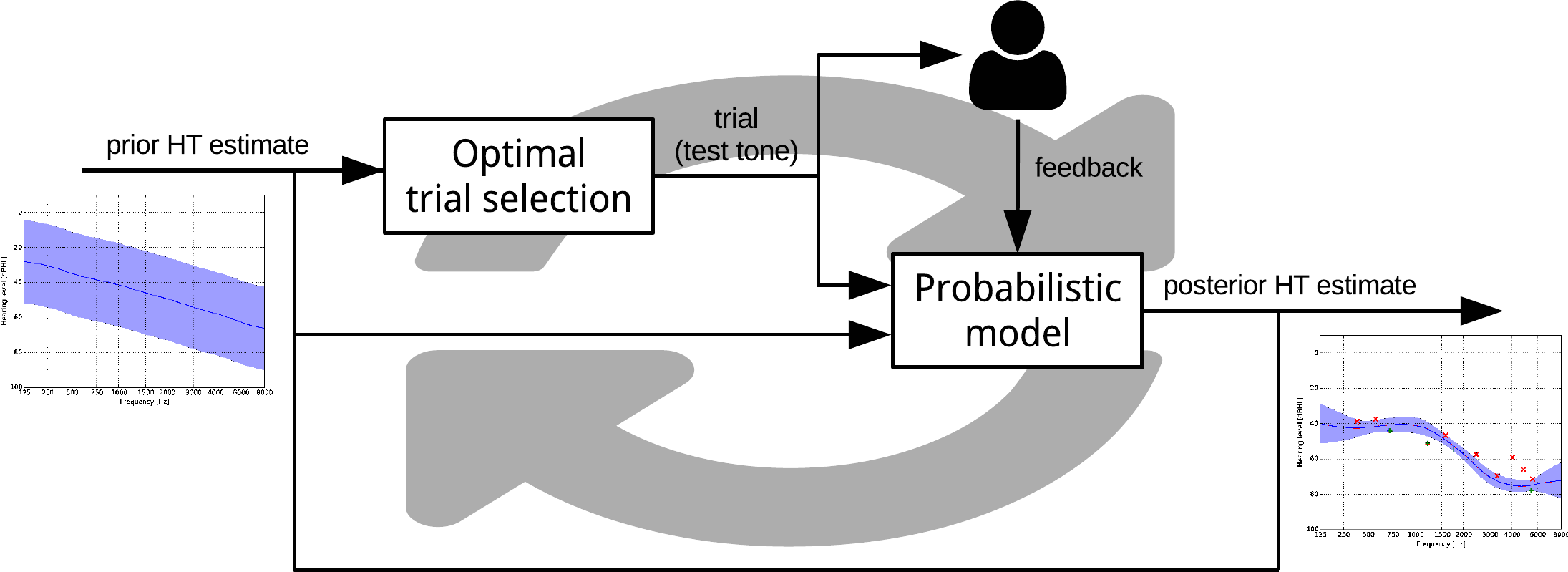}
\caption{High level overview of the proposed GP-PTA method. The probabilistic model is a two-dimensional binary classifier specified by a GP in the frequency dimension and by a psychometric function in the intensity dimension. The posterior distribution of the threshold curve is inferred incrementally by repeating two steps. First, the most informative next trial (stimulus) is determined based on the data processed so far. Next, Bayesian inference techniques are utilized to find the posterior distribution of the HT curve based on the person's response and the prior distribution.}
\label{fig:flowdiagram}
\end{figure}

This paper presents a full probabilistic approach to the pure tone hearing threshold estimation problem that addresses these drawbacks. Taking a probabilistic approach allows accounting for uncertainty in the patient's responses as well as incorporating prior knowledge in a fundamental way. Multiple probabilistic methods have already been proposed to address the first two drawbacks, for example in \citep{barthelme_flexible_2008}. The novelty of our approach is in probabilistically modelling the HT as a continuous function of frequency, rather than using multiple independent response models (psychometric functions) for a discrete set of standard frequencies. This is by endowing the threshold curve with a Gaussian process (GP) prior \citep{rasmussen_gaussian_2006}, which is why we refer to our method as `GP-PTA'. If the responses are binary (audible or non-audible), the full model can be interpreted as a two-dimensional binary classifier that is specified by a GP in the frequency dimension and by a psychometric (sigmoid) function in the intensity dimension. It provides uncertainty bands on the resulting threshold estimate, which enables fundamental and objective stopping criteria. Moreover, it allows one to find the optimal next stimulus based on the data processed so far, and stimuli at any frequency can be used. These properties reduce the number of trials required to achieve the desired accuracy. Minimizing the amount or required trials is important to reduce the cognitive burden on the (often elderly) patients. Figure \ref{fig:flowdiagram} depicts the high level structure of the GP-PTA method.

More specifically, our GP-PTA method combines the following elements:
\begin{itemize}
\item A probabilistic response model is specified to capture uncertainty in the patient's responses (Section \ref{sec:setup}). Accounting for inconsistency in the responses is crucial for providing error bands on the estimated threshold curve.
\item A generative probabilistic model is derived by combining the response model with a GP (Section \ref{sec:binclass}). The GP provides a natural way to take prior knowledge about the threshold curve into account, and relaxes the piecewise linear assumption for the threshold curve to a more natural smoothness assumption.
\item Bayesian inference techniques are applied to derive an inference algorithm for the probabilistic model. This algorithm is used to `learn' the threshold curve from the data.
\item An active learning loop is obtained by adaptively selecting the optimal next stimulus according to the ``Bayesian active learning by disagreement'' (BALD) objective function (Section \ref{sec:opt_trial_select}).
\end{itemize}
These elements are combined in simulations to evaluate the performance of the method.

\section{Experiment setup and response model}
\label{sec:setup}
To estimate a patient's pure tone HT curve, an \textit{experiment} consisting of a sequence of \textit{trials} is conducted. A trial consists of the patient labelling a pure tone stimulus with a certain frequency $f$ and intensity level as either `audible' or `non audible'. The sound intensity level is expressed by the \textit{hearing level} $h$ in dB \textit{Hearing Level} ($\dBHL$), which is defined as the intensity relative to the hearing threshold of a person with no hearing deficit ($0~\dBHL$). The patient's response is captured in binary class label $y$ according to:
\begin{equation}
    \label{eqn:classlabel}
    y(f,h)=
    \begin{cases}
        +1,  & \text{if } (f,h) \text{ is audible},\\
        -1,  & \text{otherwise}.
    \end{cases}
\end{equation}

Thus, a trial is represented by the 3-tuple $(f,h,y)$. Multiple instances of the same stimulus close to the hearing threshold will not always be labelled consistently due to human inconsistency and contextual changes. To capture this uncertainty in the responses, the `true' HT is assumed to be evaluated under white Gaussian perceptual noise: $\mathcal{N}(0,\sigma_p^2)$. This assumption leads to a probabilistic response model:
\begin{equation}
    \label{eqn:response_model}
    \begin{aligned}
    P(y|f,h) & = \text{Pr}\{y \cdot (h - \mathcal{HT}(f)) > \mathcal{N}(0,\sigma_p^2)\} \\
             & = \int_{-\infty}^{y \cdot (h-\mathcal{HT}(f))} \mathcal{N}(h'|0,\sigma_p^2) \mathrm{d}h' \\
             & = \Phi \left( \dfrac{y \cdot (h-\mathcal{HT}(f))}{\sigma_p} \right),
    \end{aligned}
\end{equation}
where $\mathcal{HT}$ is the unknown `true' hearing threshold function and $\Phi$ is the cumulative density function of the standard normal distribution. Note that $P(y|f,h)$ is a Bernoulli distribution. This response model assumes that the perceptual noise variance $\sigma_p^2$ is independent of frequency, an assumption that can easily be relaxed in the future.

\section{PTA by Bayesian binary classification}
\label{sec:binclass}
The response model from (\ref{eqn:response_model}) can be interpreted as a binary classifier for stimuli $(f,h)$. This binary classifier has a single decision boundary in the frequency-intensity space, defined by the unknown function $\mathcal{HT}(\cdot)$. Let $\mathcal{D}$ denote a data set consisting of $N$ trials:
\[
\mathcal{D} \triangleq \{(f,h,y)_1, \ldots, (f,h,y)_N\}.
\]
The goal is to use the response model to estimate $\mathcal{HT}$ as accurately as possible from $\mathcal{D}$. A good method will achieve high accuracy with a minimal number of trials. The conventional and most simple way to estimate $\mathcal{HT}$ is to restrict $f$ to a set of standard frequencies, and to estimate $\mathcal{HT}(f)$ at those frequencies by maximizing the likelihood of the data points involving that frequency. However, this approach has serious drawbacks as discussed in the introduction. To address those drawbacks, we propose a probabilistic, Bayesian approach to the estimation problem. This involves endowing the unknown function $\mathcal{HT}$ with a prior to get a full generative model for the data, and then using Bayesian inference techniques to find the posterior distribution over $\mathcal{HT}$. This approach allows processing trials at any frequency, uses all information from the trials, provides uncertainty bands on the final HT estimate, and relaxes the common piecewise linearity assumption for the HT curve to a more realistic smoothness assumption. First, we define the full generative model (Section \ref{ssec:binclass_def}) and show how to generate predictions from it. Next, we derive an inference algorithm to find the posterior distribution over $\mathcal{HT}$ (Section \ref{ssec:binclass_inference}).

\subsection{Model definition}
\label{ssec:binclass_def}
To extend the response model from (\ref{eqn:response_model}) to a full generative model, one has to specify $\mathcal{HT}$ and $\sigma_p$. The perceptual noise variance $\sigma_p^2$ is assumed to be (approximately) known in this work, but it is important to note that this quantity in principle can also be estimated by Bayesian inference.

\subsubsection{Frequency warping}
To define a prior on $\mathcal{HT}(f)$, it is convenient to first transform (warp) the frequency domain to a psychoacoustically relevant domain. Defining the prior in a psychoacoustical domain instead of directly in the frequency domain simplifies matters since the smoothness of the HT curve is more constant in the psychoacoustical domain, and our aim is to encode this smoothness property in the prior on $\mathcal{HT}$. A psychoacoustical domain is defined such that it matches the (nonlinear) human perception of frequency shifts. Multiple psychoacoustical scales are being used in practice, based on different definitions. Popular scales include the semitone scale, Mel scale, and Bark scale \citep{yost_fundamentals_1994}. The semi-logarithmic horizontal axis of an audiogram is closely related to those scales. In this work, the psychoacoustical domain is defined by the Bark scale, but it is important to note that our method in general is independent of the chosen transformation.

Multiple transformations from frequency to Bark scale have been proposed in the literature, for example in \citep{traunmuller_analytical_1990,wang_auditory_1991}. The transformations differ in optimization criteria, accuracy, and complexity. The GP-PTA model requires an invertible transformation, which is why we use the simple transformation proposed by \cite{wang_auditory_1991}:
\begin{equation}
    \label{eqn:bark}
    \bark(f) \triangleq 6*\sinh^{-1}\left(\dfrac{f}{600}\right),
\end{equation}
with $f$ in $\Hz$. We denote the transformed frequency as $x = \bark(f)$. The unknown `true' HT function in the transformed domain is denoted as
\begin{equation}
    \label{eqn:ht_bark}
    g(x) \triangleq \mathcal{HT}(\bark^{-1}(x)),
\end{equation}
and so as a consequence $\mathcal{HT}(f)=g(\bark(f))$. In the remainder of this paper, the frequency in the psychoacoustical domain, $x$, is used instead of $f$ itself.

\subsubsection{Full generative model}
The prior on $\mathcal{HT}$ is defined indirectly by the prior on $g$.
Unknown function $g$ is endowed with a Gaussian process prior, which defines a distribution over all continuous real-valued functions. A GP is fully specified by a collection of random variables, a mean function, and a covariance (kernel) function. Let
\begin{equation}
    \label{eqn:ht_gp_prior}
    g(\+x) \sim \mathcal{GP}(m(\+x), k_{\+\theta}(\+x, \+x))
\end{equation}
denote that function $g$ is drawn from a Gaussian process with mean function $m(\cdot)$ and covariance function $k_{\+\theta}(\cdot,\cdot)$. The covariance function is governed by hyperparameters $\+\theta$. The GP prior is appropriate here because it supports non-linear functions and leads to a tractable (approximate) posterior distribution. The mean and covariance functions encode our prior knowledge about $g$. In particular, we use the squared exponent covariance function, which encodes a smoothness assumption. For more background on GPs, we refer to \citep{rasmussen_gaussian_2006}.

Since $g$ is a random function, its function values can be represented by random variables. These random variables are denoted by $g_x \triangleq g(x)$ for scalar $x$, and by $\+g_{\+x} \triangleq [g(x_1),\ldots,g(x_n)]^T$ for vector $\+x$.

The full generative model is now obtained by replacing $\mathcal{HT}$ in the response model (\ref{eqn:response_model}) by latent function $g$:
\begin{subequations}
    \label{eqn:gen_model}
    \begin{align}
        P(y,g|x,h) &= P(y|g,x,h) \cdot p(g), \label{eqn:gen_model_factors} \\
        P(y|g,x,h) &= \Phi \left( \dfrac{y\cdot(h-g(x))}{\sigma_p} \right), \label{eqn:gen_model_likelihood} \\
        p(g)       &= \mathcal{GP}(m(\cdot), k_{\+\theta}(\cdot, \cdot)), \label{eqn:gen_model_gp_prior}
    \end{align}
\end{subequations}
where $p$ denotes a continuous probability distribution and $P$ a discrete one. Note that $y$ is a random variable whereas $g$ is a random function.
If $\sigma_p$ is known, the complete posterior is given by
\begin{equation}
    \label{eqn:complete_posterior}
    P(y,g|x,h;\mathcal{D}) = P(y|g,x,h) \cdot p(g|\mathcal{D}).
\end{equation}
Since the first term is independent of $\mathcal{D}$, model inference in this setting amounts to finding the posterior distribution over the latent function, $p(g|\mathcal{D})$. The posterior distribution on $\mathcal{HT}$ itself is then given by $\mathcal{HT}(f) = g(\bark(f))$, which can be evaluated for arbitrary $f$.

\subsubsection{Generating predictions}
Once the posterior distribution $p(g|\mathcal{D})$, is known, the generative model can be used to predict the patient's response to any test stimulus $(x_*,h_*)$. The posterior response distribution is given by:
\begin{equation}
    \label{eqn:classifier}
P(y_*|x_*,h_*;\mathcal{D}) = \Phi \left( \dfrac{y_*\cdot(h_*-\mu_*)}{\sqrt{\sigma_p^2+\sigma_*^2}} \right),
\end{equation}
where $\mu_*$ and $\sigma_*^2$ are the mean and variance of the (approximate) posterior GP evaluated at $x_*$. Appendix \ref{apx:predictions} contains the derivation of this result.

\subsection{Approximate model inference}
\label{ssec:binclass_inference}
A GP prior only yields a GP posterior under Gaussian likelihood \citep{rasmussen_gaussian_2006}. The likelihood $P(y|g,h)$ in the proposed model is not Gaussian, resulting in an intractable posterior $p(g|\mathcal{D})$. A common approach is to approximate the intractable posterior with a Gaussian one using Laplace's method, which we will apply here. Alternative approaches include approximation by expectation propagation, variational optimization, or Markov chain Monte Carlo (MCMC) sampling. The choice for a specific approximate inference algorithm is independent of the model definition.

The Laplace inference algorithm for the GP binary classifier is well known, and is derived for example in \citep{rasmussen_gaussian_2006}. Our model is not a pure GP classifier since the likelihood in (\ref{eqn:gen_model_likelihood}) depends on $h$, which results in a slightly different derivation. Let $\+x$, $\+h$, and $\+y$ be vectors holding the data in $\mathcal{D}$, where $\+x = \bark(\+f)$. The posterior on the latent function, $p(g|\mathcal{D})$, is completely specified by $p(\+{g_x}|\mathcal{D})$ combined with the posterior mean and covariance functions. GP inference consists of two parts: (i) approximating $p(\+{g_x}|\mathcal{D})$ with a Gaussian distribution, and (ii) finding the optimal hyperparameters for the posterior covariance function. The latter is usually done by maximizing the log-marginal likelihood of the data w.r.t. the hyperparameters, which can be applied out-of-the-box to this model. In the remainder of this section we will focus on the non-standard first part.

The true posterior $p(\+{g_x}|\+x, \+h, \+y)$ is approximated by a Gaussian distribution $q(\+{g_x}|\+x, \+h, \+y)$, which is found using the Laplace approximation:
\begin{subequations}
    \label{eqn:laplace_posterior}
    \begin{align}
    q(\+{g_x}|\+x, \+h, \+y) =& \mathcal{N}(\hat{\+g}_{\+x}, A^{-1}), \label{eqn:laplace_posterior_q} \\
     \hat{\+g}_{\+x} \triangleq& \arg \max_\+{g_x} p(\+{g_x}|\+x, \+h, \+y), \label{eqn:laplace_posterior_argmax} \\
    A \triangleq& -\nabla_{\+{g_x}}^2 \log p(\+{g_x}|\+x, \+h, \+y) |_{\+{g_x}=\hat{\+g}_{\+x}} . \label{eqn:laplace_posterior_A}
    \end{align}
\end{subequations}
The Laplace approximation reduces the inference problem to the problem of finding $\hat{\+g}_{\+x}$ and $A$, which is worked out in appendix \ref{apx:Laplace}. Once the approximate posterior has been found, it can be evaluated at any test point $x_*$ using the standard GP formula by integrating out latent function $g$:
\begin{subequations}
    \begin{align}
        q(g_*|\+x,\+h,\+y,x_*) &= \int p(g_*|g,x_*) \cdot q(g|\+x,\+h,\+y) ~\mathrm{d}g = \mathcal{N}(\mu_*, \sigma_*^2) \\
        \mu_* &= m(x_*) + k_\theta(x_*,\+x)^T k_\theta(\+x,\+x)^{-1} \hat{\+g}_{\+x},  \\
        \sigma_*^2 &=  k_\theta(x_*,x_*) - k_\theta(x_*,\+x)^T (k_\theta(\+x,\+x)+W^{-1})^{-1} k_\theta(x_*,\+x).
    \end{align}
    \label{eqn:laplace_predict}
\end{subequations}

\section{Optimal trial selection}
\label{sec:opt_trial_select}
In the setting of measuring a patient's HT, one has the ability to pick the next stimulus based on the information gathered so far. An optimal estimation strategy should not only use all available information from the experiment, but also actively select the stimuli whose responses will provide the most information about the quantities of interest. Both contribute to a lower number of trials being required, which reduces the cognitive burden on the (often elderly) person in question. The approach of interleaving optimal trial selection with learning is often referred to as ``active learning''. This section describes the derivation of a trial selection method based on the ``Bayesian Active Learning by Disagreement (BALD)'' objective proposed by \cite{houlsby_bayesian_2011}.

Given current data set $\mathcal{D}_n$ and posterior $p(g|\mathcal{D}_n)$, the goal is to find the next stimulus $(x_*,h_*)$ that will provide the most information about latent function $g$. Once the class label has been observed, the new data point is added to the data set: $\mathcal{D}_{n+1} \leftarrow \mathcal{D}_{n} \cup (x_*,h_*,y_*)$. Next, the new posterior $p(g|\mathcal{D}_{n+1})$ is approximated using the inference algorithm and the process is repeated until the uncertainty about $g$ drops below the desired threshold. Our goal is to pick the trial that ``maximizes the decrease in expected posterior entropy'' of $g$ \citep{houlsby_bayesian_2011}:
\begin{equation}
    \label{eqn:ed_obj_naive}
    (x_*,h_*) = \arg \max_{(x,h)} \Ent[g|x,h,\mathcal{D}_n] - \mathbb{E}_{y \sim P(y|x,h,\mathcal{D}_n)}\Ent[g|y,x,h,\mathcal{D}_n],
\end{equation}
where $\Ent[\cdot|\cdot]$ denotes the conditional Shannon entropy\footnote{Shannon entropy is a measure of uncertainty about the value of a random variable: $\Ent[A] \triangleq \mathbb{E}[-\log p(A)]$. Similarly, a conditional entropy relates to the entropy of a conditional probability distribution: $\Ent[A|B] \triangleq \mathbb{E}[-\log p(A|B)]$.}. The objective function from (\ref{eqn:ed_obj_naive}) is equivalent to the conditional mutual information\footnote{Mutual information is a symmetrical measure of dependence between two random variables, which can be expressed in terms of (conditional) entropies: $\text{I}[A;B]=\text{I}[B;A]=\Ent[A]-\Ent[A|B]$.} of $g$ and $y$: $\text{I}[g;y|x,h,\mathcal{D}_n]$. Due to the symmetry of mutual information, $g$ and $y$ in (\ref{eqn:ed_obj_naive}) can be swapped to get an expression that is easier to compute:
\begin{equation}
    \label{eqn:ed_obj}
    (x_*,h_*) = \arg \max_{(x,h)} \Ent[y|x,h,\mathcal{D}_n] - \mathbb{E}_{g \sim p(g|\mathcal{D}_n)}\Ent[y|g,x,h].
\end{equation}
The rewritten objective only involves conditional entropies of $y$, which is a binary random variable. Therefore, the general entropy function reduces to the binary entropy function \footnote{The binary entropy function $\ent(\cdot)$ relates the Bernoulli parameter of a binary random variable to its Shannon entropy: $A \sim \text{Bernoulli}(p) \Rightarrow \Ent[A] = \ent(p) \triangleq -p\log(p) - (1-p)\log(1-p)$.}.
The first term in (\ref{eqn:ed_obj}) is obtained by plugging in the expression for the predictive class distribution under the approximate posterior GP:
\begin{equation}
    \label{eqn:ed_obj_term1}
    \Ent[y|x,h,\mathcal{D}_n] \stackrel{(\ref{eqn:predict})}{\approx} \ent \left(  \Phi \left( \dfrac{h-\mu_x}{\sqrt{\sigma_n^2+\sigma_x^2}} \right) \right).
\end{equation}
The second term in (\ref{eqn:ed_obj}) is intractable but can be approximated very well by replacing the binary entropy by a squared exponential function as proposed in \cite{houlsby_bayesian_2011}:
\begin{equation}
    \label{eqn:ed_obj_term2}
    \begin{aligned}
    \mathbb{E}_{g \sim p(g|\mathcal{D}_n)}\ent[y|g,x,h] &\stackrel{(\ref{eqn:gen_model_likelihood})}{\approx}  \int \ent \left(  \Phi \left( \dfrac{h-g_x}{\sigma_n} \right) \right) \mathcal{N}(g_x|\mu_x,\sigma_x^2) \mathrm{d}g_x \\
     &~\approx \int \exp \left( - \dfrac{(h-g_x)^2}{\sigma_n^2 \pi \ln 2} \right) \mathcal{N}(g_x|\mu_x,\sigma_x^2) \mathrm{d}g_x \\
     &~= \dfrac{C}{\sqrt{\sigma_x^2 + C^2}} \exp \left( \dfrac{-(h-\mu_x)^2}{2(\sigma_x^2 + C^2)} \right),
    \end{aligned}
\end{equation}
where $C = \sigma_n \sqrt{\frac{\pi \ln 2}{2}}$.
In both terms, the first approximation includes the fact that we approximate the `true' intractable posterior distribution of $g$ by a GP. Substituting both terms in (\ref{eqn:ed_obj}) yields the complete expression for the BALD optimum:
\begin{equation}
    \label{eqn:ed_obj_total}
    (x_*,h_*) = \arg \max_{(x,h)} \ent \left(  \Phi \left( \dfrac{h-\mu_x}{\sqrt{\sigma_n^2+\sigma_x^2}} \right) \right) - \dfrac{C}{\sqrt{\sigma_x^2 + C^2}} \exp \left( \dfrac{-(h-\mu_x)^2}{2(\sigma_x^2 + C^2)} \right),
\end{equation}
where $(x,h)$ should of course be constrained to the domain of interest.

For fixed $x$, the objective function is maximized by setting $h = \mu_x$ (the decision boundary of the classifier). The proof of this property is contained in appendix \ref{apx:BALD_h_proof}. This result implies that the optimization problem reduces to a one dimensional one:
\begin{equation}
    \label{eqn:ed_obj_total_x}
    x_* = \arg \min_{x} \dfrac{C}{\sqrt{\sigma_x^2 + C^2}} = \arg \max_{x} \sigma_x^2.
\end{equation}

So maximizing the BALD objective in this case results in picking the frequency for which the posterior GP variance is largest. However, if the assumption of the response noise being independent of frequency is relaxed, this will no longer be the case. Unfortunately, the objective function is not differentiable w.r.t. $x$ due to the approximate GP inference method that (non-linearly) relates $\sigma_x$ to $x$. However, evaluating the objective function is cheap since it only involves an evaluation of the pre-computed approximate posterior GP using (\ref{eqn:laplace_predict}). Since $x$ is one dimensional, a simple line search works well in practice.

Experiments show that the BALD objective results in frequently performing trials at the borders of the frequency range. Although this is functionally correct, it is not desirable from an audiological point of view. To reflect the fact that the accuracy of the HT estimate is less important at very low and very high frequencies, the objective function may be weighted by a frequency dependent factor.

\section{Related work}
\label{sec:related}
Existing pure tone hearing threshold estimation methods range from the simple empirical staircase scheme \citep{carhart1959preferred} to more theory-based methods such as QUEST \citep{watson1983quest} and CAST \citep{ozdamar_classification_1990}. An adaptive Bayesian method for estimating the parameters of a flexible psychometric function is described in \citep{barthelme_flexible_2008}. In \citep{de2010efficient}, a Gaussian mixture model is used to model the HT curve and to take prior knowledge into account. The main difference between these existing methods and our GP-PTA method is that the existing ones all focus on estimating the parameters of some psychometric function at a fixed set of frequencies, and ignore the correlation between these parameters at different frequencies. In the context of our framework, this would correspond to estimating the parameters of multiple independent response models: one for every frequency of interest. A recent poster titled ``Optimizing Pure-Tone Audiometry Using Machine Learning'' \citep{barbour_optimizing_2015} describes a method similar to GP-PTA, in which the continuous HT curve is also modelled by a Gaussian process. Due to the limited amount of details, we are unable to determine the exact amount of similarity between their work and ours.

\section{Simulations}
\label{sec:results}
The response model, inference method, and trial selection method are combined to incrementally estimate an unknown hearing threshold curve. The approach is tested in simulations using fictional but representative HT curves. To obtain a reasonable prior for the threshold curve, we use a set of standard audiograms \citep{bisgaard_standard_2010}. More specifically, a linear prior with fixed variance is obtained by performing linear regression in the warped frequency domain on the average audiogram resulting from the set of standard audiograms. The simple squared exponential kernel is used in the Gaussian process prior to encode the smoothness assumption on the threshold curve. The kernel hyperparameters are optimized after every trial by maximizing the approximate log marginal likelihood w.r.t. the hyperparameters.

\begin{figure}[!ht]
\centering
\subfloat[Audiogram based on prior]{
  \includegraphics[width=.5\textwidth]{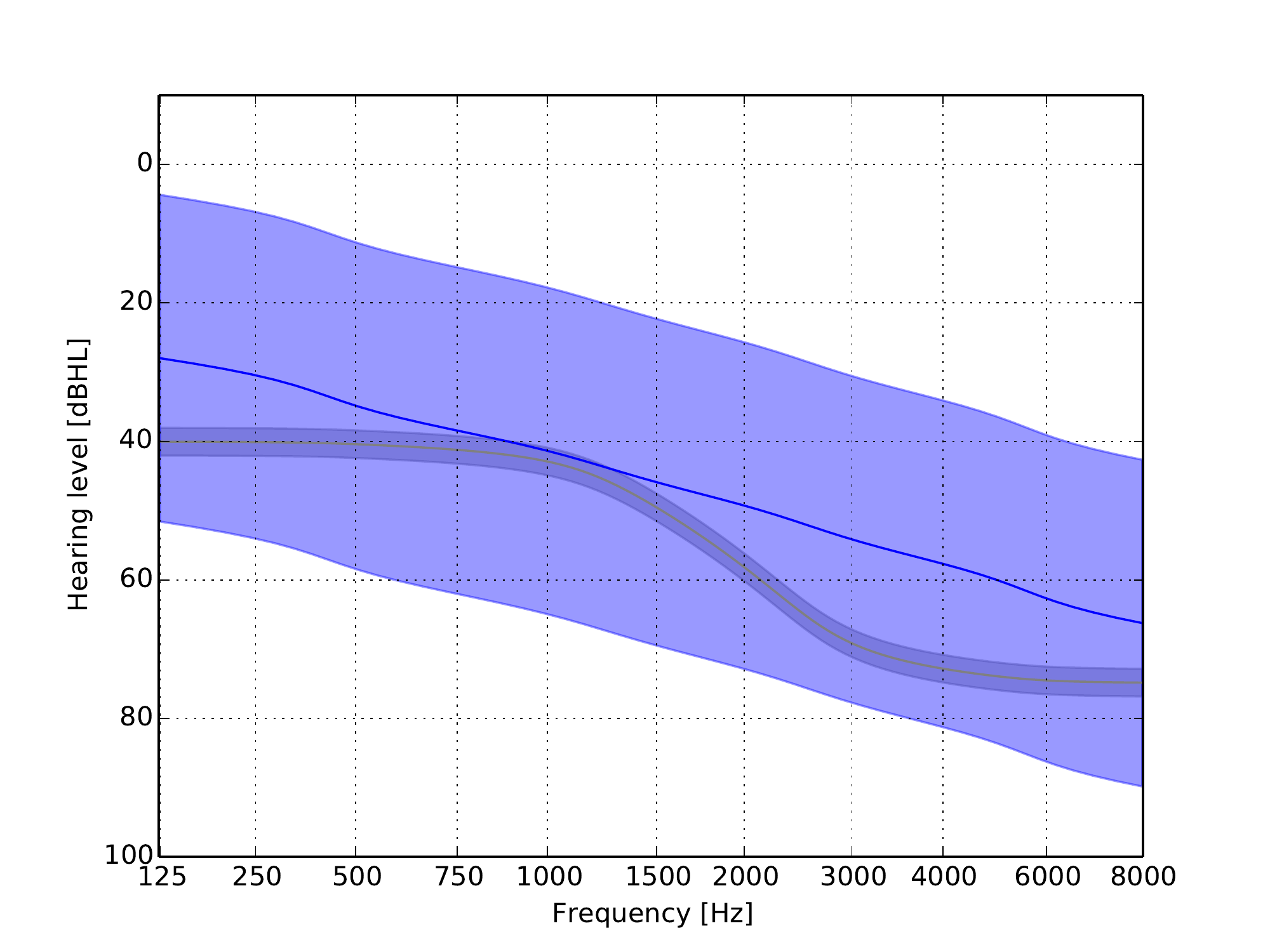}
}
\subfloat[Audiogram after 7 trials]{
  \includegraphics[width=.5\textwidth]{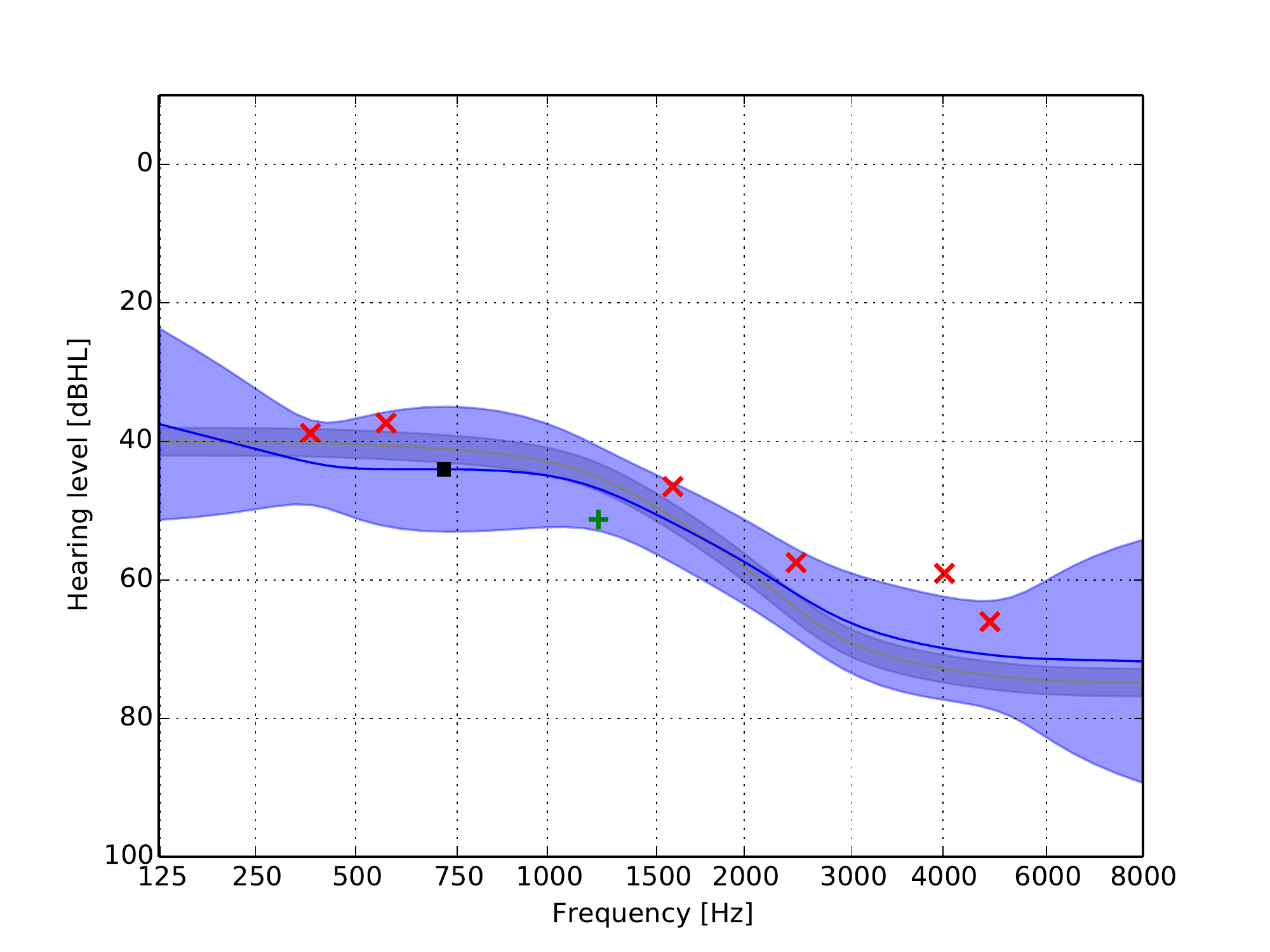}
}
\hspace{0mm}
\subfloat[Audiogram after 14 trials]{
  \includegraphics[width=.5\textwidth]{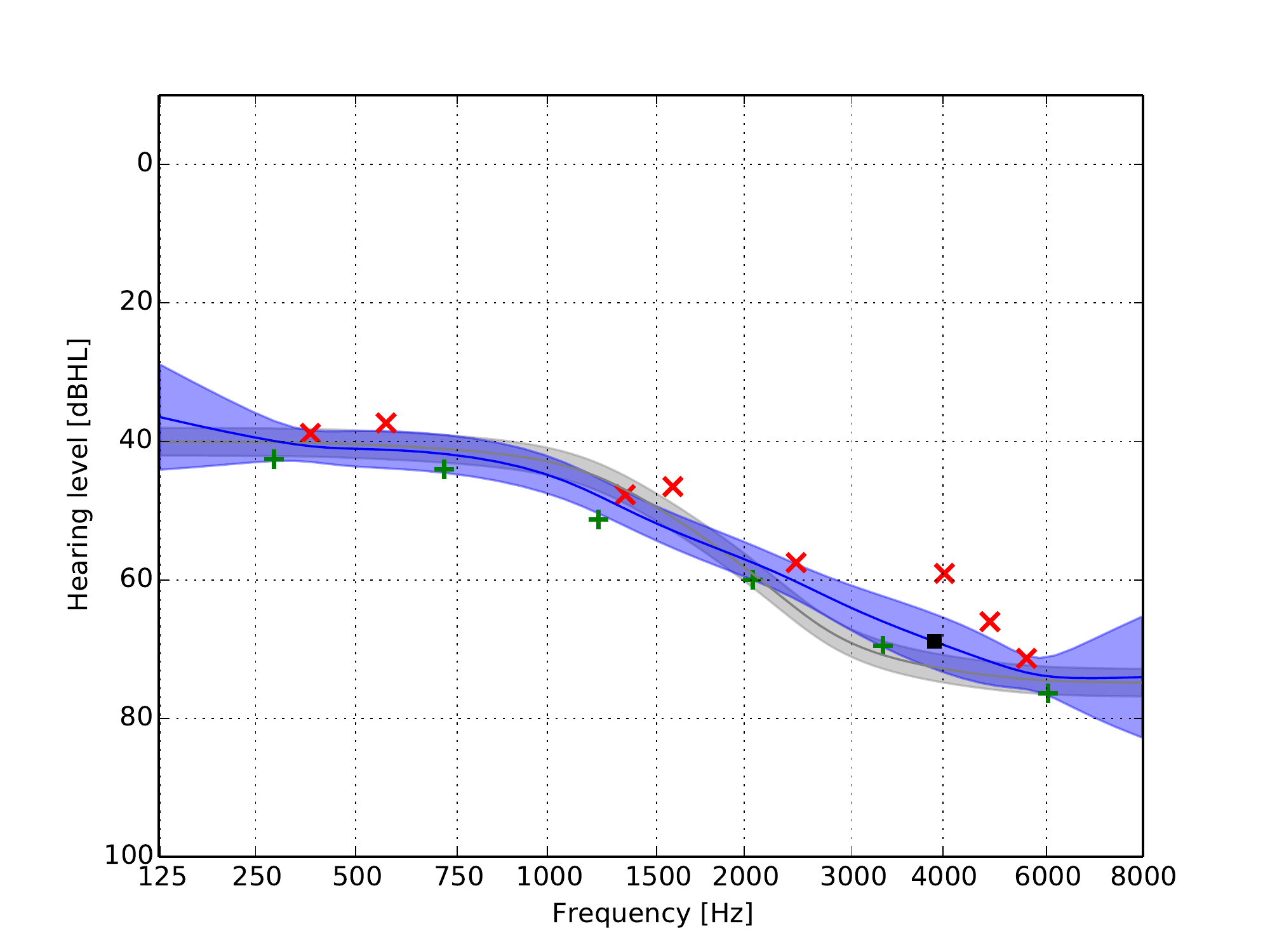}
}
\subfloat[Audiogram after 21 trials]{
  \includegraphics[width=.5\textwidth]{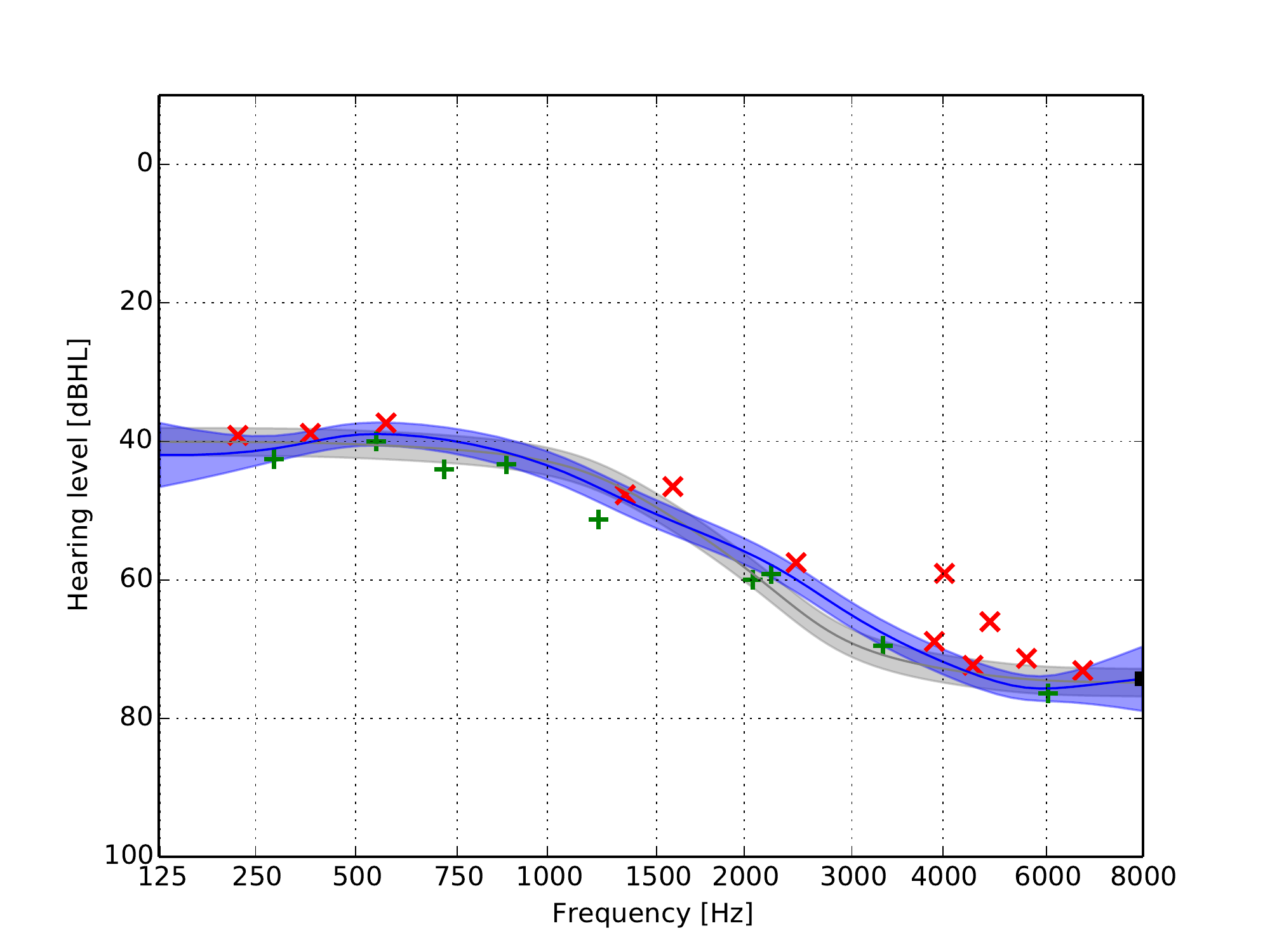}
}
\caption{Incremental estimation of a fictional but representative hearing threshold curve with fixed response noise variance. The shaded areas represent a single standard deviation. The pluses and crosses indicate audible and non audible stimuli, respectively. The square indicates the proposed next trial.}

\label{fig:simulation1}
\end{figure}

\begin{figure}[!ht]
	\centering
	\includegraphics[width=.7\textwidth]{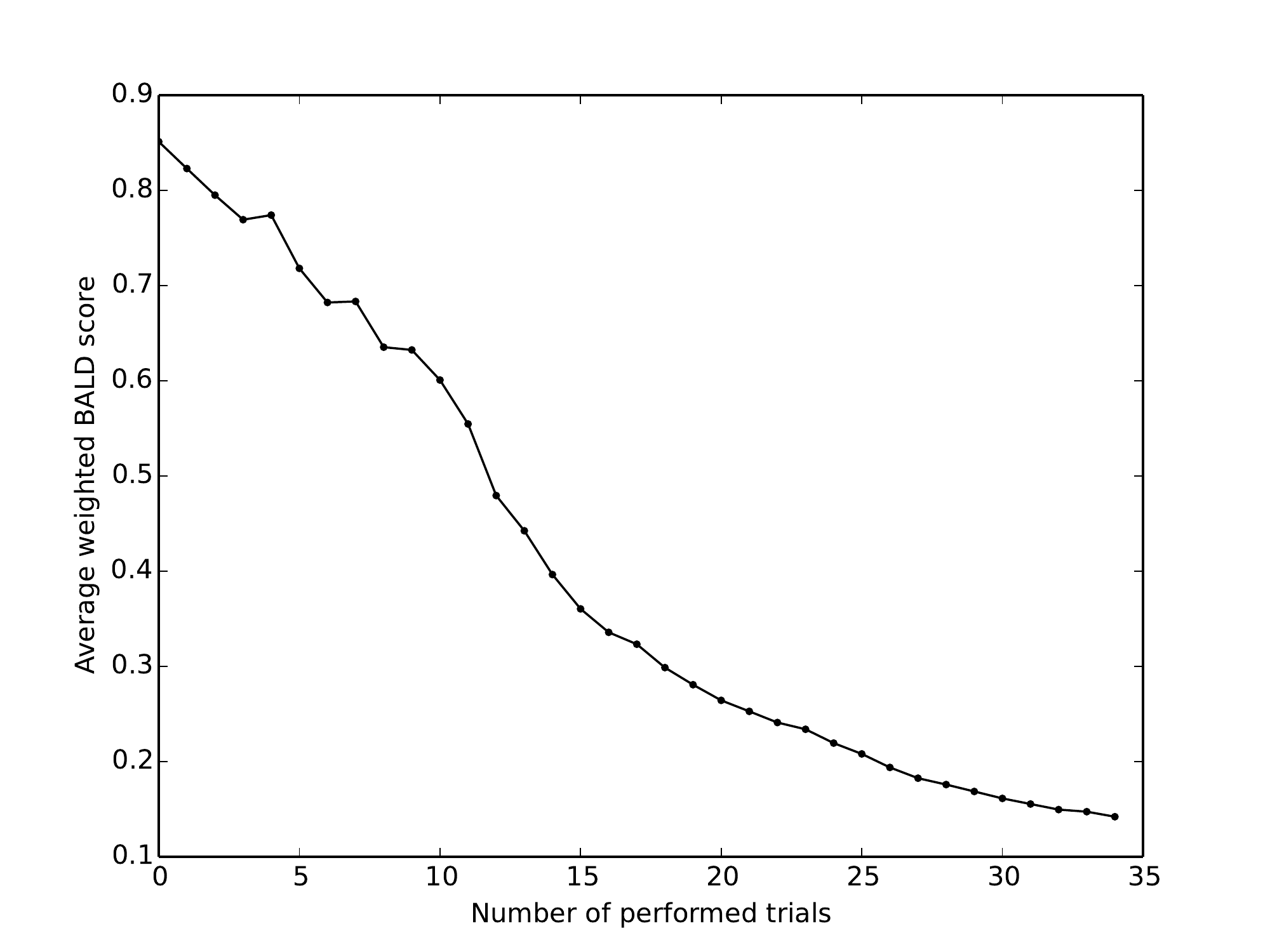}
	\caption{Evolution of the mean of the trial selection objective function (BALD).}
	\label{fig:simulation1_bald}
\end{figure}

Figure \ref{fig:simulation1} depicts the incremental estimation of a HT curve under response noise standard deviation of $2~\dBHL$. As is to be expected, increasing the response noise variance leads to slower convergence of the estimate due to more `false' class labels. Figure \ref{fig:simulation1_bald} depicts the evolution of the mean value of the BALD objective function over time. Ideally, this curve should be monotonically decreasing since every trial contains at least some information. However, the fact that the hyperparameters are optimized using maximum likelihood estimation can cause the mean BALD score to increase on some occasions.

\section{Discussion}
\label{sec:discussion}
Taking a probabilistic approach to the problem of estimating the pure tone hearing threshold curve has the important advantage that all forms of uncertainty can be accounted for in a fundamental and objective way. This removes the need for empirically optimized but rather `ad-hoc' approaches like ``up $5~\dB$ - down $10~\dB$''. Moreover, it provides an optimal strategy for selecting the next trial. By accounting for the uncertainty in the patient's responses and in the prior knowledge about the threshold curve, uncertainty bands on the estimated threshold curve can be obtained in a fundamental way. Taking uncertainty in the responses into account is also important from a psychological point of view. Confidence in the correctness of the final estimate should increase if the user can rely on the estimation method to handle answers like ``I think I heard it, but I'm not completely sure'' are handled in a robust and objective way. The GP-PTA method described in this paper assumes binary answers, and all uncertainty about the response is captured in the response model. However, it is easy to extend the response model to allow more subtle answers, for example by introducing a neutral option. Real-life experiments are required to investigate the usefulness of such extensions. The sigmoid function in our response model takes the role of psychometric function, and follows directly from the the perceptual noise assumptions. Instead of fine-tuning the parameters of the psychometric function itself, as is described for example in \citep{barthelme_flexible_2008}, we capture the uncertainty about the psychometric function in the prior on the threshold curve.

We show that a GP prior on the threshold curve is practical, since it provides a distribution over all possible continuous threshold curves. Moreover, it combines a realistic smoothness assumption with a posterior that can can be approximated analytically. The assumption that the prior perceptual noise variance is known is not unrealistic, since it can be derived from (large) data sets like we did for the simulations. The assumption that this variance is independent of frequency might not be as realistic. However, it is straightforward to extend the model to support frequency-dependent perceptual noise variance. In that case, the BALD criterion does not simplify to picking trials at the frequency with the highest uncertainty about the threshold estimate, which is why we include the complete derivation.

The proposed GP-PTA method provides automatic procedures for selecting the best next trial as well as updating the estimated pure-tone hearing threshold on the basis of patient trial response. As a result, our method in principle supports automated fitting of dynamic range compressing circuits in hearing aids without need for professional human assistance. Moreover, the uncertainty bands on the fitted parameters might be exploited by in-situ tuning algorithms for the hearing aid. Automating these processes is important to be able to cost-effectively cope with the number of hearing impaired persons that is growing worldwide at an alarming rate \citep{kochkin2005marketrak,lin2011hearing}. According to the world health organization, the majority of the 360 million people with disabling hearing loss live in low- and middle-income countries, and the current production of hearing aids meets less than 10\% of the global need \citep{who_factsheet_fs300}.

\section*{Acknowledgements}
We like to thank Thijs van de Laar and Tjalling Tjalkens for the interesting discussions, and the GN ReSound research department for scientific support.

\newpage

\begin{appendices}
%\titleformat{\section}{\large\bfseries}{APPENDIX~\thesection :}{0.5em}{}

\section{Generating predictions}
\label{apx:predictions}
The complete posterior (\ref{eqn:complete_posterior}) of the generative model is converted into a binary classifier $(x,h) \rightarrow y$ by marginalizing out latent function $g$. If the posterior Gaussian process $p(g|\mathcal{D})$ is known, the posterior response distribution for a test stimulus $(x_*,h_*)$ is given by:
\begin{equation}
    \label{eqn:predict}
    \begin{aligned}
    P(y_*|x_*,h_*;\mathcal{D})      &=~~~\int P(y_*,g|x_*,h_*;\mathcal{D}) ~\mathrm{d}g \\
                                    &\!\stackrel{(\ref{eqn:gen_model})}{=}~~~  \int \Phi \left( \dfrac{y_*\cdot(h_*-g(x_*))}{\sigma_p} \right) \cdot p(g|\mathcal{D}) ~\mathrm{d}g \\
                                    &=~~~ \int \Phi \left( \dfrac{y_*\cdot(h_*-g_*)}{\sigma_p} \right) \cdot \left[ \int p(g_*|g,x_*) \cdot p(g|\mathcal{D}) ~\mathrm{d}g \right] ~\mathrm{d}g_* \\
                                    &\!\!\!\!\!\! \stackrel{\text{Laplace}}{\approx} \int \Phi \left( \dfrac{y_*\cdot(h_*-g_*)}{\sigma_p} \right) \cdot \mathcal{N}(g_*|\mu_*,\sigma_*^2) ~\mathrm{d}g_* \\
                                    &=~~~ \Phi \left( \dfrac{y_*\cdot(h_*-\mu_*)}{\sqrt{\sigma_p^2+\sigma_*^2}} \right),
    \end{aligned}
\end{equation}
where $g_*$ denotes the latent function value $g(x_*)$. In general, the above integral is not analytically tractable since $p(g|\mathcal{D})$ has no exact analytical solution due to the non-Gaussian likelihood. In this paper, the intractable (non-GP) posterior is approximated by a GP using the Laplace approximation. Under this approximation, the posterior distribution of $g_*$ is Gaussian, and the integral can be solved analytically (see Section 3.9 of \cite{rasmussen_gaussian_2006} for the detailed derivation).

\section{Laplace approximation of the posterior GP}
\label{apx:Laplace}
Laplace's method can be applied to find a Gaussian approximation to an intractable distribution. In this case, the intractable posterior $p(\+{g_x}|\+x, \+h, \+y)$ is approximated by a multivariate Gaussian distribution $q(\+{g_x}|\+x, \+h, \+y)$ according to:
\begin{subequations}
    \label{eqn:laplace_posterior_apx}
    \begin{align}
    q(\+{g_x}|\+x, \+h, \+y) =& \mathcal{N}(\hat{\+g}_{\+x}, A^{-1}), \label{eqn:laplace_posterior_q_apx} \\
     \hat{\+g}_{\+x} \triangleq& \arg \max_\+{g_x} p(\+{g_x}|\+x, \+h, \+y), \label{eqn:laplace_posterior_argmax_apx} \\
    A \triangleq& -\nabla_{\+{g_x}}^2 \log p(\+{g_x}|\+x, \+h, \+y) |_{\+{g_x}=\hat{\+g}_{\+x}} . \label{eqn:laplace_posterior_A_apx}
    \end{align}
\end{subequations}

We will derive an algorithm for finding $\hat{\+g}_{\+x}$ and $A$. Applying Bayes' rule yields:
\begin{equation}
\label{eqn:laplace_bayes}
p(\+{g_x}|\+x, \+h, \+y) = \frac{p(\+y|\+h,\+{g_x})p(\+{g_x}|\+x)}{p(\+y|\+x,\+h)},
\end{equation}
where the denominator is independent of $\+{g_x}$.
Since prior $p(g)$ is a GP, $p(\+{g_x}|\+x)$ is Gaussian. Without loss of generality, a zero mean function is assumed for simplicity (the mean function can always be absorbed in the likelihood function):
\begin{equation}
    \label{eqn:laplace_gp_prior}
    p(\+{g_x}|\+x) = \mathcal{N}(\+0, k_\theta(\+x,\+x)) = \mathcal{N}(\+0, K_{\+x\+x}).
\end{equation}
Taking the logarithm of the numerator of (\ref{eqn:laplace_bayes}) and writing out the logarithm of the Gaussian prior results in:
\begin{equation}
    \begin{array} {lcl}
    \Psi(\+{g_x}) & \triangleq & \log p(\+y|\+h,\+{g_x}) + \log p(\+{g_x}|\+x) \\
    &=& \log p(\+y|\+h,\+{g_x}) - \frac{1}{2} \+{g_x}^T K_{\+x\+x}^{-1}\+{g_x} - \frac{1}{2} \log |K_{\+x\+x}| - \frac{N}{2} \log 2\pi.
    \end{array}
    \label{eqn:laplace_psi}
\end{equation}
To find $\hat{\+g}_{\+x}$ and $A$, the first and second order derivatives of $\Psi$ w.r.t. $\+{g_x}$ are required:
\begin{subequations}
    \begin{align}
    \nabla_{\+{g_x}} \Psi(\+{g_x}) = & \nabla_{\+{g_x}} \log p(\+y|\+h,\+{g_x}) - K_{\+x\+x}^{-1}\+{g_x}, \\
    \nabla_{\+{g_x}}^2 \Psi(\+{g_x}) = & \nabla_{\+{g_x}}^2 \log p(\+y|\+h,\+{g_x}) - K_{\+x\+x}^{-1} = -W - K_{\+x\+x}^{-1},
    \end{align}
    \label{eqn:laplace_psi_derivatives}
\end{subequations}
where $W$ is a diagonal matrix since the likelihood factors over the data points. The log-likelihood is defined by the response model from (\ref{eqn:gen_model_likelihood}):
\begin{equation}
    \label{eqn:laplace_log_lik}
    \log p(\+y|\+h,\+{g_x}) = \log \Phi \left( \dfrac{(\+h-\+{g_x})\+y}{\sigma_p} \right).
\end{equation}
The first and second order derivatives of the log likelihood w.r.t. $\+{g_x}$ are required in (\ref{eqn:laplace_psi_derivatives}), and we state them here without derivation:
\begin{subequations}
    \begin{align}
    \nabla_{\+{g_x}} \log p(\+y|\+h,\+{g_x}) &= \frac{\+y\mathcal{N}(\+{g_x}|\+h,\Sigma_p)}{\Phi \left(\frac{(\+{g_x} - \+h)\+y}{\sigma_p}\right)}, \\
    \nabla_{\+{g_x}}^2 \log p(\+y|\+h,\+{g_x}) &= \frac{- \mathcal{N}(\+{g_x}|\+h,\Sigma_n)^2}{\Phi \left(\frac{(\+{g_x} - \+h)\+y}{\sigma_n}\right)^2} - \frac{\+y(\+{g_x}-\+h)\mathcal{N}(\+{g_x}|\+h,\Sigma_n)}{\Phi \left(\frac{(\+{g_x} - \+h)\+y}{\sigma_n}\right)},
    \end{align}
    \label{eqn:laplace_log_likelihood_derivatives}
\end{subequations}
where $\Sigma_n = \sigma_n^2 I_N$, a $N \times N$ diagonal matrix. Finally, $\hat{\+g}_{\+x}$ is found by setting the first order derivative of $\Psi$ to zero:
\begin{equation}
    \nabla_{\+{g_x}} \Psi(\hat{\+g}_{\+x}) = \+0 \Longrightarrow \hat{\+g}_{\+x} = K_{\+x\+x} \cdot \nabla_{\+{g_x}}  \log p(\+y|\+h,\+{g_x}).
    \label{eqn:laplace_g_hat}
\end{equation}
This non-linear equation can be solved by a Newton-Rhapson algorithm. Analog to the derivation in Section 3.4.1 of \citep{rasmussen_gaussian_2006}, the Newton steps are given by:
\begin{equation}
    \label{eqn:laplace_newton_update}
    \begin{array} {lcl}
    \+{g_x}^\text{new} &=& \+{g_x} - (\nabla_{\+{g_x}}^2 \Psi)^{-1} \cdot \nabla_{\+{g_x}} \Psi \\
    & = & \+{g_x} + (K_{\+x,\+x}^{-1}+W)^{-1} \cdot [ \nabla_{\+{g_x}} \log p(\+y|\+h,\+{g_x}) - K_{\+x,\+x}^{-1} \+{g_x}] \\
    & = & (K_{\+x,\+x}^{-1}+W)^{-1} [W\+{g_x} + \nabla_{\+{g_x}} \log p(\+y|\+h,\+{g_x})].
    \end{array}
\end{equation}
Repeating (\ref{eqn:laplace_newton_update}) until convergence of $\+{g_x}$ yields $\hat{\+g}_{\+x}$. The total Laplace approximation is obtained by substituting the results in the definition of (\ref{eqn:laplace_posterior_apx}):
\begin{equation}
    \label{eqn:laplace_total}
    p(\+{g_x}|\+x, \+h, \+y) \approx q(\+{g_x}|\+x, \+h, \+y) = \mathcal{N}(\hat{\+g}_{\+x}, (K_{\+x,\+x}^{-1}+W)^{-1}).
\end{equation}

\section{Proof of BALD optimum at $h=\mu_x$}
\label{apx:BALD_h_proof}
\begin{trm:bald_optimum_h}
The BALD objective function
\[
\mathrm{BALD}(x,h) =  \ent \left(  \Phi \left( \dfrac{h-\mu_x}{\sqrt{\sigma_n^2+\sigma_x^2}} \right) \right) - \dfrac{C}{\sqrt{\sigma_x^2 + C^2}} \exp \left( \dfrac{-(h-\mu_x)^2}{2(\sigma_x^2 + C^2)} \right)
\]
is maximized w.r.t. $h$ by $h = \mu_x$. Constant $C$ is given by $C = \sigma_n \sqrt{\frac{\pi \ln 2}{2}}$, $\sigma_n \geq 0$, and $\sigma_x \geq 0$.
\end{trm:bald_optimum_h}
\begin{proof}
We approximate the binary entropy term up to $\mathcal{O}(h^4)$ by a squared exponential function as proposed in \cite{houlsby_bayesian_2011}:
\[
\ent \left(  \Phi \left( \dfrac{h-\mu_x}{\sqrt{\sigma_n^2+\sigma_x^2}} \right) \right) \approx \exp \left( \dfrac{-(h-\mu_x)^2}{\pi \ln2(\sigma_n^2+\sigma_x^2)} \right).
\]
Substituting this approximation yields the approximate (smooth and differentiable) objective function $\Psi$:
\[
\mathrm{BALD}(x,h) \approx \Psi(x,h) \triangleq \exp \left( \dfrac{-(h-\mu_x)^2}{\pi \ln2 (\sigma_n^2+\sigma_x^2)} \right) - \dfrac{C}{\sqrt{\sigma_x^2 + C^2}} \exp \left( \dfrac{-(h-\mu_x)^2}{2(\sigma_x^2 + C^2)} \right).
\]
To find the maxima, we differentiate $\Psi$ w.r.t. to $h$:
\[
\begin{aligned}
\dfrac{\partial \Psi}{\partial h} = & \dfrac{-2}{\pi\ln2(\sigma_n^2+\sigma_x^2)} \cdot (h-\mu_x) \cdot \exp \left( \dfrac{-(h-\mu_x)^2}{\pi \ln2(\sigma_n^2+\sigma_x^2)} \right) \\
 & + \dfrac{2\sigma_n \sqrt{\pi\ln2}}{(\pi\ln2\sigma_n^2+2\sigma_x^2)^{\frac{3}{2}}} \cdot (h-\mu_x) \cdot \exp \left( \dfrac{-(h-\mu_x)^2}{\pi \ln2\sigma_n^2+2\sigma_x^2} \right),
\end{aligned}
\]
where constant $C$ has been written out. It is easily verified that $\frac{\partial \Psi}{\partial h}\vert_{h=\mu_x} = 0$ and that $\Psi$ has a local maximum there. We are left with the problem of proving that this is the global maximum. Another extreme point $h_* \neq \mu_x$ will have to satisfy $\frac{\partial \Psi}{\partial h}\vert_{h=h_*} = 0$.
\[
\dfrac{\partial \Psi}{\partial h} \bigg\vert_{h = h_* \in \mathbb{R} \backslash \mu_x} = 0 \Rightarrow \dfrac{1}{(h_*-\mu_x)} \dfrac{\partial \Psi}{\partial h} \bigg\vert_{h = h_*} = 0.
\]
Taking the logarithm and simplifying yields:
\[
(h_*-\mu_x)^2 \underbrace{\left[ \dfrac{1}{\pi \ln2(\sigma_n^2+\sigma_x^2)} - \dfrac{1}{\pi \ln2 \sigma_n^2 + 2\sigma_x^2} \right]}_{P \leq 0} = \underbrace{ \ln\left[ \dfrac{2 (\pi \ln2\sigma_n^2+ 2\sigma_x^2)^\frac{3}{2} }{\pi \ln2(\sigma_n^2+\sigma_x^2) 2\sigma_n\sqrt{\pi\ln2}} \right] }_{Q},
\]
which is a second order polynomial in $(h_*-\mu_x)$. Since coefficient $P$ is negative, the equality can only be satisfied if $Q < 0$, a necessary condition for the existence of another extreme point of $\Psi$. Simplifying $Q$ yields:
\[
\begin{aligned}
Q =& \ln(2) + \frac{3}{2}\ln(\pi \ln2\sigma_n^2 + 2\sigma_x^2) - \frac{3}{2}\ln(\pi \ln2) - \ln(2) - \ln(\sigma_n) - \ln(\sigma_n^2+\sigma_x^2) \\
  =& \frac{3}{2}\ln\left(1+\frac{2}{\pi\ln2}\frac{\sigma_x^2}{\sigma_n^2} \right) - \ln\left(1+\frac{\sigma_x^2}{\sigma_n^2} \right) \\
  =& \frac{3}{2}\ln\left(1+\frac{2}{\pi\ln2}u \right) - \ln\left(1+u \right),
\end{aligned}
\]
where we used the substitution $u \triangleq \frac{\sigma_x^2}{\sigma_n^2} \geq 0$. It is immediately clear that $Q=0$ in case $u=0$. For the case $u>0$ we turn to the first order derivative of $Q$ w.r.t. $u$:
\[
\begin{aligned}
\frac{\partial Q}{\partial u} = & \frac{3}{2} \cdot \frac{1}{1+\frac{2}{\pi\ln2}u} \cdot \frac{2}{\pi\ln2} - \frac{1}{1+u} \\
 =& \frac{u+3-\pi\ln2}{2u^2+(\pi\ln2+2)u+\pi\ln2} > 0,
\end{aligned}
\]
where we used the observation that both the numerator and denominator are positive for $u \geq 0$. Combined with the fact that $Q=0$ in case $u=0$, it follows that $Q \geq 0$, which violates the necessary condition for the existence of another extreme point of $\Psi$. The consequence is that $\Psi$ has just one maximum for fixed $x$, at $h = \mu_x$.
\end{proof}

\end{appendices}

\newpage

% References
\bibliographystyle{apalike}

\begin{thebibliography}{}

\bibitem[Barbour, 2015]{barbour_optimizing_2015}
Barbour, D. (\textbf{2015}).
\newblock ``Optimizing {Pure}-{Tone} {Audiometry} {Using} {Gaussian} {Processes},''
\newblock Proc. 38th Annual MidWinter Meeting, Association for Research
  in Otolarynology, p. 38.

\bibitem[Barthelm\'{e} and Mamassian, 2008]{barthelme_flexible_2008}
Barthelm\'{e}, S. and Mamassian, P. (\textbf{2008}).
\newblock ``A flexible {Bayesian} method for adaptive measurement in psychophysics,''
\newblock {\tt {arXiv}:0809.0387 [stat]}.

\bibitem[Bisgaard et~al., 2010]{bisgaard_standard_2010}
Bisgaard, N., Vlaming, M. S. M.~G., and Dahlquist, M. (\textbf{2010}).
\newblock ``Standard audiograms for the {IEC} 60118-15 measurement procedure,''
\newblock {Trends in Amplification} 14, 113--120.

\bibitem[Carhart and Jerger, 1959]{carhart1959preferred}
Carhart, R. and Jerger, J. (\textbf{1959}).
\newblock ``Preferred method for clinical determination of pure-tone thresholds,''
\newblock {J. of Speech \& Hearing Disorders} 24, 330--345.

\bibitem[de~Vries et~al., 2010]{de2010efficient}
de~Vries, A., Stadler, S., Leijon, A., Dijkstra, T., and Ypma, A. (\textbf{2010}).
\newblock ``Efficient evaluation of hearing ability,''
\newblock US Patent App. 12/429,783 (7 October 2010).

\bibitem[Houlsby et~al., 2011]{houlsby_bayesian_2011}
Houlsby, N., Husz\'{a}r, F., Ghahramani, Z., and Lengyel, M. (\textbf{2011}).
\newblock ``Bayesian active learning for classification and preference learning,''
\newblock {\tt {arXiv}:1112.5745 [cs, stat]}.

\bibitem[Kochkin, 2005]{kochkin2005marketrak}
Kochkin, S. (\textbf{2005}).
\newblock ``{MarkeTrak VII}: Hearing loss population tops 31 million people,''
\newblock {The Hearing Review} 12(7), 16--29.

\bibitem[Lin et~al., 2011]{lin2011hearing}
Lin, F.~R., Niparko, J.~K., and Ferrucci, L. (\textbf{2011}).
\newblock ``Hearing loss prevalence in the {United States},''
\newblock {Archives of Internal Medicine} 171(20), 1851--1853.

\bibitem[\"{O}zdamar et~al., 1990]{ozdamar_classification_1990}
\"{O}zdamar, \"{O}., Eilers, R.~E., Miskiel, E., and Widen, J. (\textbf{1990}).
\newblock ``Classification of audiograms by sequential testing using a dynamic
  {Bayesian} procedure,''
\newblock {J. Acoust. Soc. Am.} 88(5), 2171--2179.

\bibitem[Rasmussen and Williams, 2006]{rasmussen_gaussian_2006}
Rasmussen, C.~E. and Williams, C. K.~I. (\textbf{2006}).
\newblock {\em Gaussian processes for machine learning} ({MIT} Press, Cambridge), 248 pp.

\bibitem[Traunm\"{u}ller, 1990]{traunmuller_analytical_1990}
Traunm\"{u}ller, H. (\textbf{1990}).
\newblock ``Analytical expressions for the tonotopic sensory scale,''
\newblock {J. Acoust. Soc. Am.} 88(1), 97--100.

\bibitem[Wang et~al., 1991]{wang_auditory_1991}
Wang, S., Sekey, A., and Gersho, A. (\textbf{1991}).
\newblock ``Auditory distortion measure for speech coding,''
\newblock {Proc. 1991 Int. Conf. on Acoustics, Speech, and Signal Processing (ICASSP-91)}, 493--496.

\bibitem[Watson and Pelli, 1983]{watson1983quest}
Watson, A.~B. and Pelli, D.~G. (\textbf{1983}).
\newblock ``Quest: A {Bayesian} adaptive psychometric method,''
\newblock {Perception \& psychophysics} 33(2), 113--120.

\bibitem[{World~Health~Organization}, 2015]{who_factsheet_fs300}
{World~Health~Organization} (\textbf{2015}).
\newblock {\em Fact sheet. Deafness and hearing loss.}
\newblock Available at {\em http://www.who.int/mediacentre/factsheets/fs300/en/} (accessed on July 17, 2015).

\bibitem[Yost, 1994]{yost_fundamentals_1994}
Yost, W.~A. (\textbf{1994}).
\newblock {\em Fundamentals of hearing: An introduction (3rd ed.), Vol. XIII} ({Academic Press, San Diego}), 326 pp.

\end{thebibliography}

\end{document}